\documentclass[
twocolumn, 
aps,
]{revtex4-1}

\usepackage{amssymb}
\usepackage{graphicx}
\usepackage{color,soul}
\usepackage{dcolumn}

\newcommand{\ud}{\uparrow\downarrow}
\newcommand{\uu}{\uparrow\uparrow}
\newcommand{\uuu}{\uparrow\uparrow\uparrow}
\newcommand{\uud}{\uparrow\uparrow\downarrow}
\newcommand{\udu}{\uparrow\downarrow\uparrow}

\begin{document}

\title[Role of layer interactions in  CrI$_3$-graphene MTJ's]{Role of quantum confinement and interlayer coupling in CrI$_3$-graphene magnetic tunnel junctions}

\author{Jonathan J.~Heath$^\dagger$}
\author{Marcio Costa$^{\mathsection}$}
\author{Marco Buongiorno-Nardelli$^\bot$}
\author{Marcelo A.~Kuroda$^\dagger$}%
\email{mkuroda@auburn.edu}
\affiliation{$^\dagger$ Department of Physics, Auburn University, Auburn AL 36849, USA\\
$^\mathsection$ Brazilian Nanotechnology National Laboratory (LNNano,CNPEM), Campinas, SP 13083-970, Brazil, Center for Natural Sciences and Humanities, Federal University of ABC, Santo Andr{\'e}, SP, Brazil\\
$^\bot$
Department of Physics and Chemistry, University of North Texas, Denton, TX 76203, USA}

\date{\today}


\begin{abstract}
Recent demonstrations of magnetic ordering and spin transport in two-dimensional heterostructures have opened research venues in these material systems.  In order to control and enhance the related physical phenomena, quantitative descriptions linking experimental observations to atomic details must be produced.  Here we combine first principles and quantum ballistic transport calculations to shed important insights from an atomistic viewpoint on the underlying mechanisms governing spin transport in graphene/CrI$_3$ junctions.  Descriptions of the electronic structure reveal that tunneling is the dominant transport mechanism in these heterostructures and help differentiate intermediate metamagnetic states present in the switching process. We find that quantum confinement and layer-layer interactions are key to describing transport in these two-dimensional systems.  Ballistic transport calculations further support these findings and yield magnetoresistance values in remarkable agreement with experiments. The short width of these barriers limits analysis solely based on the bulk complex band structure often employed in the description of magnetic tunnel junctions.  Our work devises mechanisms to attain larger tunneling magnetoresistances, proving valuable to the advancement of spin valves in layered heterostructures.
\end{abstract}



\maketitle


\section{Introduction}
The existence of magnetic order in two-dimensional (2D) material systems -- especially as inherited from their bulk counterparts -- was an open question until recent years. Early work by Mermin and Wagner \cite{mermin1966, bruno2001} based on an isotropic Heisenberg model with long-range order proved that at $T>0$ spontaneous magnetism and antiferromagnetism should not exist in 2D systems.  However, this idea has been challenged by theoretical analyses using first-principles calculations predicting 2D ferromagnetic semiconductors from the exfoliation of K$_2$CuF$_4$ or Cr$X$Te$_3$ ($X = $ Si, Ge) crystals \cite{sachs2013, li2014a}. Subsequent experiments in CrGeTe$_3$ \cite{tian2016, lee2016, gong2017} and CrI$_3$ \cite{huang2017} have confirmed the attainment of spin order (ferromagnetic or antiferromagnetic) in 2D materials systems.  These achievements enable fundamental studies of the impact of dimensionality on strongly correlated phenomena and offer highly desirable building blocks for spintronic devices based on 2D materials.  In very recent works, the magnetic ground-state and interlayer coupling have been probed in graphene/CrI$_3$ heterostructures forming magnetic tunnel junctions (MTJ's) \cite{klein2018} and CrGeTe$_6$ devices in field-effect transistor (FET) configurations \cite{wang2018}. At low temperatures, magneto-optical Kerr effect microscopy demonstrated electrostatic gate control of magnetism in CrI$_3$ bilayers \cite{huang2018}.  Moreover, magnetoresistances of up to 550\% were reported by switching metamagnetic states in CrI$_3$ via external magnetic fields \cite{klein2018}.

In these promising experiments, nonetheless, a quantitative link between the heterostructure composition and experimental observations is still missing. For instance, previous calculations fail to explain the observed tunneling-dominated regime as the CrI$_3$ majority bands cross the graphite leads' Fermi energy \cite{klein2018}.  Furthermore, spin filter models proposed to describe these experiments treat layers as independent and cannot capture quantum confinement effects \cite{latil2006, splendiani2010}. Recent work using density functional theory (DFT) calculations has shown that 3,000\% tunneling magnetoresistance can be attained in CrI$_3$ based tunnel junctions using Cu leads \cite{paudel2019spin}. However, comprehensive studies examining the composition and magnetic configuration dependence of spin transport properties in these junctions are still missing. For example, the small Fermi surface of graphene leads and quantum confinement are expected to alter these responses. Hence, atomistic descriptions accounting for layer interactions, electrodes, and external fields are required to not only understand and control the mechanisms governing spin transport in these systems \cite{butler2001spin, karpan2007graphite}, but also to design complex heterostructures based on 2D materials \cite{geim2013van} exploiting the spin degree of freedom.

Here we describe the tunneling magnetoresistance (TMR) in graphene/CrI$_3$/graphene heterostructures using first principles calculations within the DFT and Landauer's formalism for ballistic transport. Our results reveal that tunneling is the dominant transport mechanism, reconciling atomistic descriptions to experimental observations \cite{klein2018}.  Analysis of the band structure of few layer CrI$_3$ junctions reveals that the interplay between quantum confinement and interactions between layers is essential to defining band alignments and the resulting tunneling barriers.  As a consequence, the effective spin tunneling barriers vary with both thickness and magnetic ordering in the CrI$_3$ layers.  The magnetoresistance values obtained by employing Landauer's formalism to various metamagnetic states in bilayer and trilayer junctions exhibit quantitative agreement with available experimental results \cite{klein2018}.  We also discuss  limitations in the use of bulk CrI$_3$ complex band structures to gauge tunneling rates in these ultrathin junctions. Outcomes of this work highlight the importance of weak interactions and atomistic details in these layered magnetic systems which may be exploited towards the development of future magnetic tunnel junctions.


\section{Methods}
\begin{figure}[htpb]
\centering{
\includegraphics[width=3.2in]{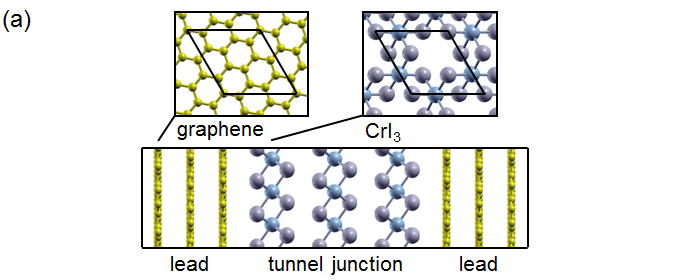} \\
\includegraphics[width=3.5in]{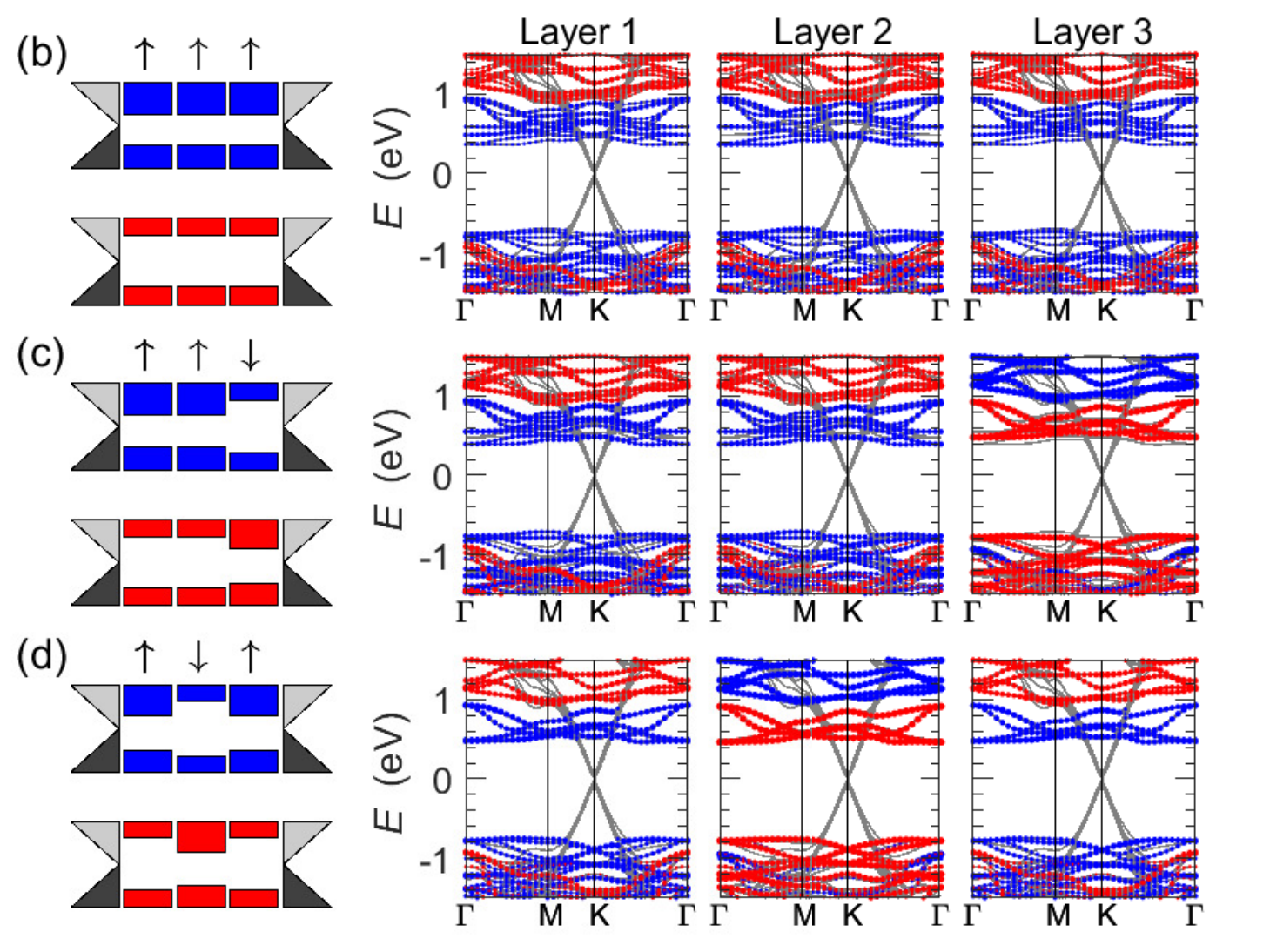}
\caption{(a) Schematic of the supercells: ($1\times1$)-CrI$_3$/($\sqrt{7}\times\sqrt{7}$)-graphene leads. Solid lines in the different cross-sections mark the unit cell of each layer.  Band diagrams and corresponding band structures in graphene/trilayer CrI$_3$/graphene junctions for various metamagnetic states: (b) $\uuu$ state; (c) $\uud$ state; and (d) $\udu$ state. In each subpanel, Bloch states are projected onto localized atomic orbitals of different CrI$_3$ layers in the MTJ, allowing the band alignments to be identified for the  spin majority (blue) and minority (red) populations. In all cases, graphene electrodes preserve the linear dispersion of the monolayer and the Fermi level resides within the band gap, demonstrating that tunneling transport is the dominant mechanism \cite{supp_mat}}}
\label{fig:bandstruc}
\end{figure}

The systems considered here are formed by CrI$_3$ junctions and graphene electrodes. The supercells representing the tunnel junctions consist of up to three CrI$_3$ layers arranged by $abc$ stacking \cite{mcguire2015}. The epitaxy of these cells accommodates $1\times1$ CrI$_3$ layers on $\sqrt{7}\times\sqrt{7}$ graphene, where the in-plane lattice constant is set to that of the CrI$_3$ ($a \approx 6.79$~\AA), yielding a 4\% lateral tensile strain applied to the graphene electrodes. Three graphene layers on each side of the junction serve as the leads for the heterostructure, as illustrated in Fig.~\ref{fig:bandstruc}(a). Additionally, periodic boundary conditions are assumed in all three dimensions. In order to diminish thickness dependent dispersion found in Bernal stacking \cite{latil2006, latil2007} and avoid band splitting, graphene supercells are turbostratically stacked $ab_{\theta}$ ($\theta$ = 21.79$^o$) to form highly oriented graphene electrodes \cite{shallcross2010electronic, kim2017tunable, bao2011stacking}. The equilibrium configurations for the CrI$_3$/graphene heterojunctions are obtained by holding the in-plane lattice constant $a$ fixed while allowing the out-of-plane lattice parameter $c$ of the entire supercell and all atoms to fully relax. The resulting interlayer distances between adjacent CrI$_3$ layers are approximately 3.35~\AA~regardless of the system's metamagnetic state while CrI$_3$/graphene separations are around 3.54~\AA, in good agreement with other studies \cite{paudel2019spin, zhang2018strong}.


First principles calculations are performed within the DFT where the exchange-correlation energy is parameterized by spin polarized generalized gradient approximation (GGA) functionals \cite{Perdew1997}, including dispersion forces (vdW-DF-C09) \cite{Dion2004, thonhauser2007, Cooper2010}. The 2D Brillouin zone (2D-BZ) is sampled using a $6\times6$ Monkhorst-Pack mesh \cite{monkhorst1976}.  
These calculations use PAW pseudopotentials \cite{blochl1994, dalcorso2014} for the description of the atomic cores with cutoff energies of 50~Ry and 400~Ry for the Kohn-Sham wave functions and densities, respectively.  Additionally, projected density of states (DOS) and band structures of the systems are computed to analyze the results. All DFT calculations are carried out by employing the Quantum Espresso software suite \cite{Giannozzi2009, Choi1999, Smogunov2004}.  Quantum transport calculations are performed using the PAOtransport code \cite{nardelli1999, buongiorno2018} which allows efficient sampling of the 2D-BZ by producing tight-binding  models from projection of plane-wave pseudopotential wave-functions onto atomic orbitals.

\section{Results and discussion}
Results for the 2D band structures of trilayer CrI$_3$/graphene heterojunctions with different metamagnetic states (Fig.~\ref{fig:bandstruc}) indicate that tunneling is the governing transport mechanism in the magnetic junctions.  In all cases, the Dirac cone is easily identifiable, residing at the supercell's K-point for this epitaxy.  The Fermi level -- located at the tip of the Dirac cone -- lies between the conduction and valence bands.  These results, also found for monolayer and bilayer junctions \cite{supp_mat}, differ considerably with respect to those previously reported \cite{klein2018}, where the Fermi level resided at the spin majority conduction band.  We attribute these differences to the large 9\% compressive strain in the graphene layers employed in those calculations and, to a lesser extent, the description of the dispersion forces \cite{choi2010effects, zhang2018strong, supp_mat}.

While the coupling between layers does not form gap states at the graphene/CrI$_3$ interface, it is sufficient to alter field-modulated transport.  We exemplify this using the trilayer CrI$_3$ case that offers three different metamagnetic states ($\uuu$, $\uud$ and $\udu$) which yield  more diverse tunneling resistances than configurations in the bilayer junctions ($\uu$ and $\ud$). To visualize these different states [Fig.~\ref{fig:bandstruc}(b)-(d)], their band structures are computed and decomposed onto atomic orbitals localized on different magnetic layers and colored according to their spin population (blue, majority; red, minority).
In the case of the  parallel trilayer configuration ($\uuu$), the obtained band gap for the spin majority (minority) is approximately 0.05~eV (0.02~eV) smaller than the parallel bilayer configuration and 0.17~eV (0.07~eV) smaller than the monolayer.  Similar to other layered systems \cite{splendiani2010, castellanos2014}, we find that band gaps depend on thickness although not as strongly as in transition metal dichalcogenides.

Notwithstanding the weak thickness dependence, external magnetic fields allow for the modulation of band alignments, as depicted in Fig.~\ref{fig:bandstruc}. For large magnetic fields, interlayer coupling is strengthened when the magnetization of all CrI$_3$ layers is parallel, forming smaller band gaps (tunneling barriers) than in cases where the magnetization of adjacent layers is opposite. As magnetic fields diminish and one of the layers flips its magnetization, band gaps exhibit an increase, which varies depending on the metamagnetic state.  For instance, the conduction band edge of the CrI$_3$ spin majority population in the $\uud$ configuration is sensibly closer to the Fermi level than in the $\udu$ case as a result of the stronger magnetic coupling between the first two layers \cite{jang2019microscopic}. Interlayer coupling, albeit weak, precludes junctions with the same total magnetization ($|M|  = 6\mu_B$ per layer) but different magnetic ordering from being treated as equivalent spin barriers (e.g. $T_{\uud} \neq T_{\udu}$), as illustrated by schematics found in Fig.~\ref{fig:bandstruc}(b)-(d).

In order to characterize quantum transport in these junctions, the conductance is computed from the transmission probability of states in the graphene leads through the scattering region containing the 2D magnetic junction.  A priori, transport across junctions is ballistic and occurs mainly via tunneling \cite{condon1978} as thermionic-field emission \cite{szebook, padovani1966} appears to be negligible. Due to experimental conditions corresponding to the spin transport measurements ($T\lesssim 4\mbox{ K}$ and high-quality junctions), contributions from phonons or spin-flip mechanisms are omitted. In this context, zero-bias conductance per unit area (which accounts for the present metal electrodes) is computed from the transmission probability following Landauer's formalism \cite{datta1995, dicarlo1994, kuroda2011b}:
\begin{equation}
\sigma_m = \frac{G_0}{A} \sum_{s=\uparrow,\downarrow} T_s^m(E)   \label{eq:conductance}.
\end{equation}
Here $G_0 = e^2/h$ and $A$ are the conductance quantum \cite{datta1995} and the supercell's cross-sectional area, respectively, giving $G_0/A \approx 97$ S/$\mu$m$^2$.  The transmission probability  $T_s^m(E)$ corresponds to a spin channel $s$ ($\uparrow$ or $\downarrow$) when the system is in the metamagnetic state $m$. This probability is an average of the momentum-resolved transmissions $t_s^m(\mathbf{k}_{\parallel},E)$ over the 2D-BZ:
\begin{equation}
T_s^m(E)  \equiv \frac{-A}{(2\pi)^2} \int dE \int_{{\mathrm{\small{2D-BZ}}}} \,d^2\mathbf{k}_{\parallel}\,\, t_s^m(\mathbf{k}_{\parallel},E) \frac{df}{dE} \label{eq:transmission},
\end{equation}
 where $f(E)$ is the Fermi-Dirac distribution function and   $t_s^m(\mathbf{k}_{\parallel},E) = \sum_{i,j} t_{i,j,s}^m(\mathbf{k}_{\parallel},E)$ which includes contributions from all possible bands in one electrode to those in the other electrode, preserving spin $s$ and crystal momentum $\mathbf{k_{\parallel}}$.
Note that thermionic emission of either electrons or holes as well as tunneling or hopping mechanisms are implicit in Eq.~(\ref{eq:transmission}).

\begin{figure}[htpb]
\centering{
\includegraphics[width= 3.5in]{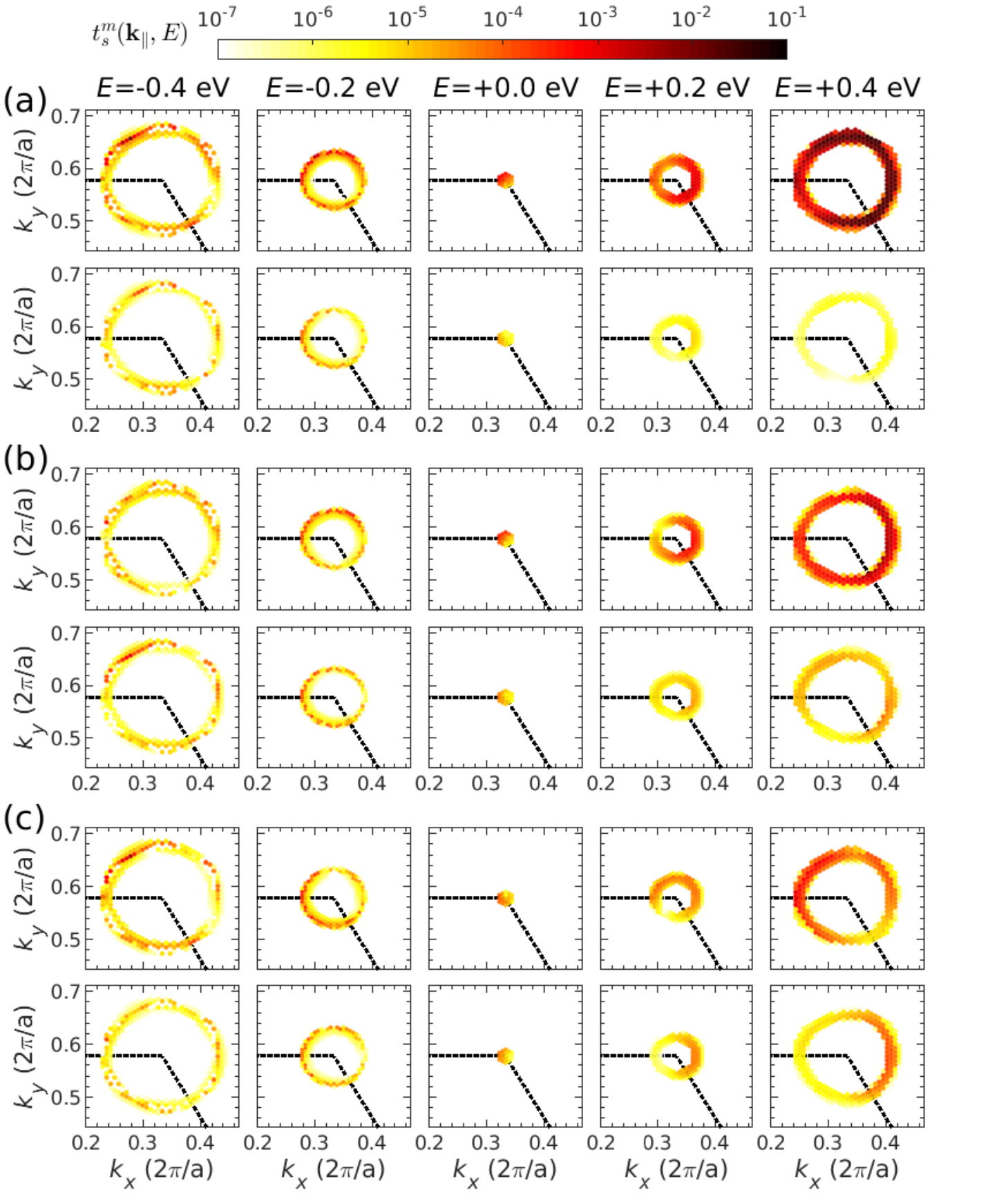}
\caption{Spin and momentum-resolved transmission profiles [$t_s^m(\mathbf{k}_\parallel,E)$] near the supercell K-point for trilayer CrI$_3$ between graphene leads for several metamagnetic states: (a) $\uuu$; (b) $\uud$; and (c) $\udu$. In each set of plots, the top (bottom) row corresponds to the spin majority (minority) channel and columns (from left to right) correspond to energies $E =$ -0.4, -0.2, 0.0, 0.2 and 0.4~eV. The dashed line denotes a corner of the hexagonal Brillouin zone.  Logarithmic scale is provided on top.}}
\label{fig:trans}
\end{figure}

\begin{figure*}[t]
\centering{
\includegraphics[width= 6.5in]{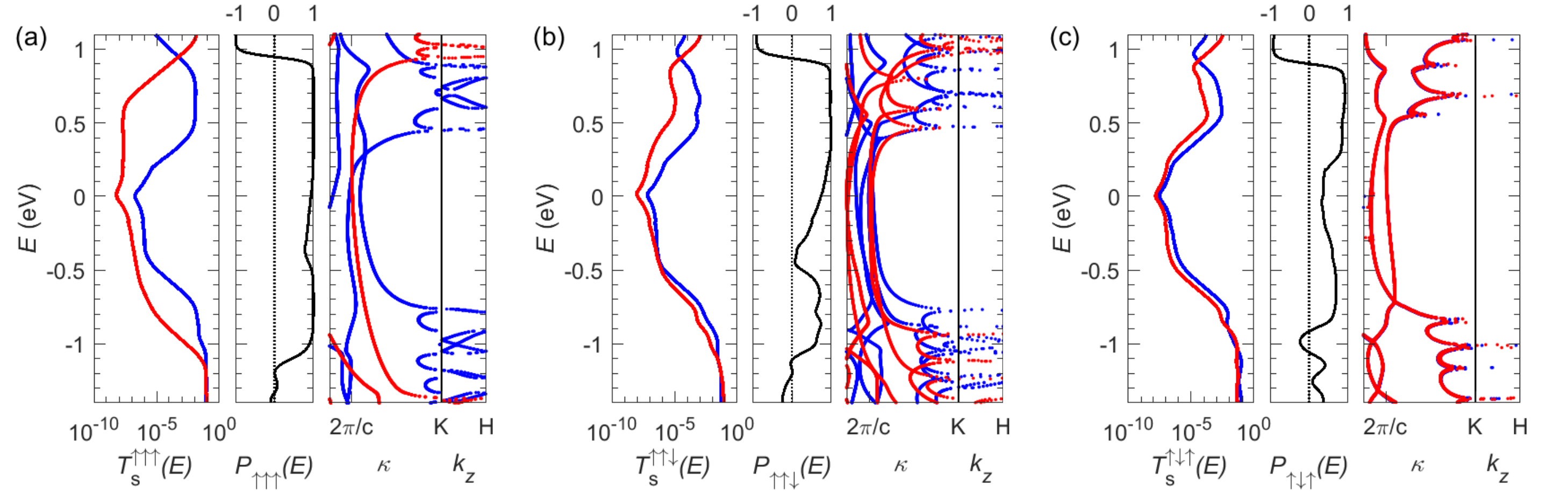}
\caption{Transport analysis of trilayer CrI$_3$ tunneling junctions in different metamagnetic states:  (a) $\uuu$, (b) $\uud$, and (c) $\udu$.  For each case, we plot (from left to right) the transmission $T_s^m(E)$ in logarithmic scale; the polarization of the tunneling current; and the bulk CrI$_3$ CBS along the tunneling direction for both spin majority (blue) and minority (red) carriers.}}
\label{fig:tmr}
\end{figure*}

Tunneling current in these systems is a combination of available states in the electrodes and their tunneling probability.  In Fig.~\ref{fig:trans}, we plot the spin and momentum resolved transmission probability $t_s^m(\mathbf{k}_\parallel,E)$ for the trilayer CrI$_3$ in metamagnetic states ($\uuu$, $\uud$, and $\udu$) that produce distinct tunneling currents.  It is important to note that due to the peculiar dispersion of graphene, contributions to transport  around the Fermi energy originate in a small portion of reciprocal space near the K-point. These transport calculations are then sampled using a finer $\mathbf{k}_\parallel$-grid (120$\times$120) \cite{nardelli1999, buongiorno2018}.  In all cases, this low density of states in the electrodes unpropitiously yields low conductance values. Net transmissions increase as the Fermi level moves away from the Dirac point due to the augmented Fermi contour. This enhancement is consistent with recent demonstrations using gated devices \cite{wang2018very, song2018giant, song2019voltage} and is asymmetric because the CrI$_3$ conduction bands reside closer to the Fermi level than the valence bands.

At fixed energies, transmission profiles show a strong dependence on the metamagnetic state $m$, even if they have the same total magnetization.  The $\uuu$ case shows the strongest (weakest) transmission for the spin majority (minority).  As one of the layers flips its magnetization, the transmissions of spin populations change differently depending on the location of the layer with opposite magnetization.  The $\uud$-configuration exhibits a larger transmission than the $\udu$ case due to smaller spin majority barriers (band gaps) produced in the first two layers through the enhanced coupling between the adjacent layers [Fig.~\ref{fig:bandstruc}(c)].

The net transmissions $T_s^m$ through trilayer CrI$_3$ as a function of the electron energy for non-equivalent metamagnetic configurations $m$ are produced in Fig.~\ref{fig:tmr}. In all cases, transmissions for both spin up (blue) and down (red) channels exhibit a dip near the Fermi level. This reduction, due to the vanishing DOS in graphene, is overcome when carrier energies move toward the CrI$_3$ band edges. Consistent with the profiles in Fig.~\ref{fig:trans}, the difference between the spin majority and minority tunneling probabilities shrinks when more adjacent layers have opposite magnetization.

For a given metamagnetic state $m$, we also compute the spin polarization of the tunneling current as:
\begin{equation}
P_m = \frac{T_{\uparrow}^m-T_{\downarrow}^m}{T_{\uparrow}^m+T_{\downarrow}^m}\label{eq:pol}.
\end{equation} Our results in Fig.~\ref{fig:tmr} also show that high spin polarization in the tunneling current can be attained. Near the Fermi level and within the band gap, the strong polarization of the spin current through the parallel configuration $\uuu$ ($P_{\uuu} \approx 0.94$) diminishes as one of the layers flips its magnetization ($P_{\uud} \approx 0.73$ and $P_{\udu} \approx 0.35$). We note that current in these last two configurations remains polarized due to the odd number of CrI$_3$ layers. A high spin polarization is also attained for the parallel bilayer configuration ($P_{\uparrow\uparrow} \approx 0.92$) and the monolayer ($P_{\uparrow} \approx 0.79$)\cite{supp_mat}.

In MTJ's, the energy dependent decay rates of evanescent wave functions can be estimated from the junction's complex band structure (CBS) \cite{butler2001spin, karpan2007graphite}.  For the different magnetic configurations of the trilayer case, we compute the bulk CrI$_3$ CBS along the K-H direction (Fig.~\ref{fig:tmr}) for spin majority and minority populations. To facilitate comparison, band alignments reflect those obtained in the graphene/CrI$_3$/graphene band structures shown in Fig.~\ref{fig:bandstruc}. We notice that the slowest decaying evanescent states exhibit weak energy dependence within the band gap and away from the band edges. Moreover, channel transport estimates based only on these evanescent modes yield transmissions about two orders of magnitude greater than those obtained from full transport calculations and exhibit no dip at the Fermi level \cite{supp_mat}.  Limitations in the use of CBS to estimate decay rates are attributed to quantum confinement and coupling to the electrodes \cite{sivadas2018stacking} in these ultrathin junctions, pointing out the importance of interactions between the leads and the junction.

\begin{figure}[htpb]
\centering{
\includegraphics[width=3.25in]{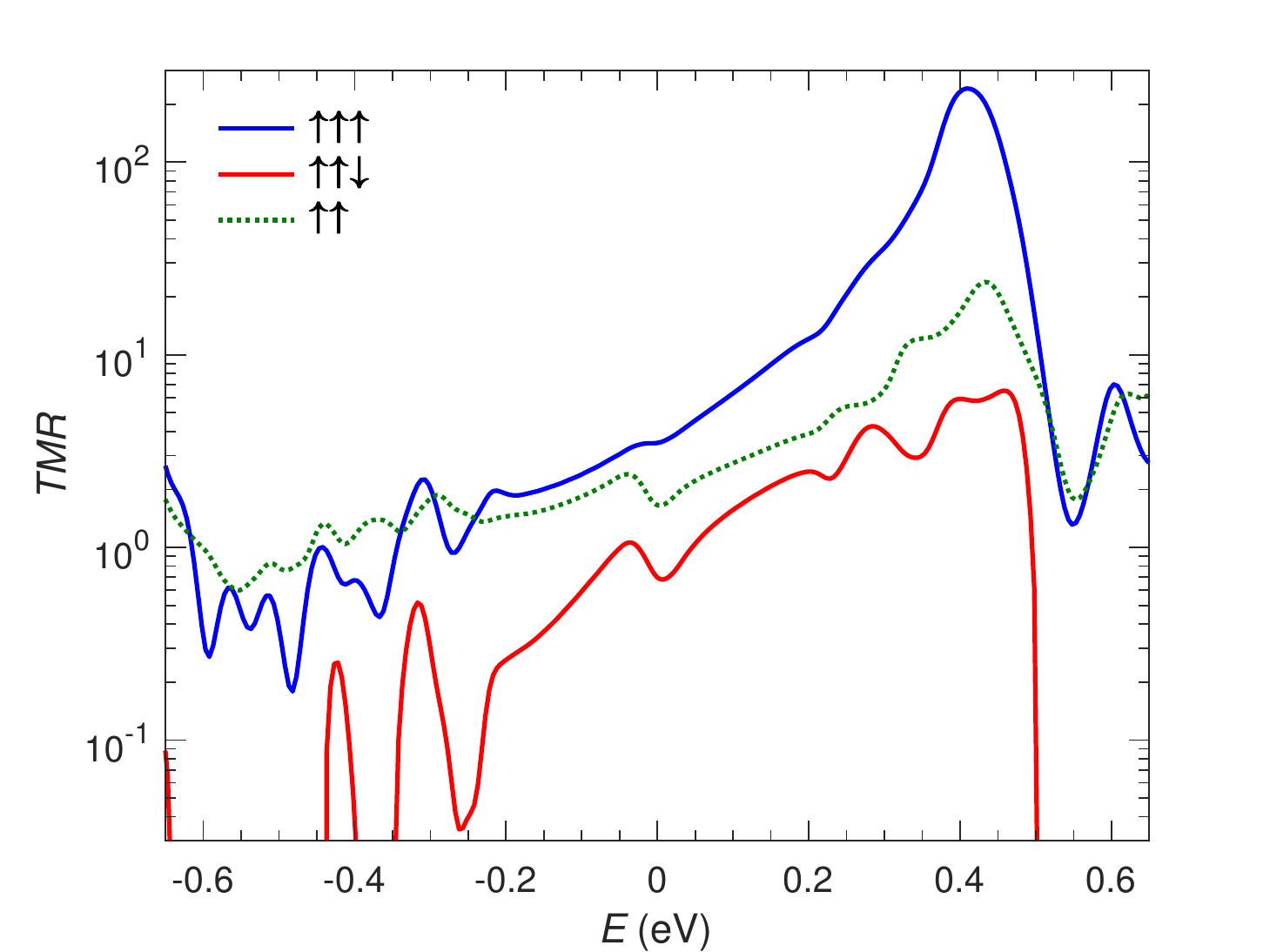}
\caption{Tunneling magnetoresistances  [Eq.~(\ref{eq:tmr})] as a function of Fermi level for both trilayer (TMR$_{\uuu}$) and (TMR$_{\uud}$) and bilayer (TMR$_{\uu}$) systems.}}
\label{fig:tmr2}
\end{figure}

The tunneling magnetoresistance (TMR) for an arbitrary metamagnetic state $m$ is defined: 
\begin{equation}
\mbox{TMR}_m = \frac{(T_{\uparrow}^m+T_{\downarrow}^m)-(T_{\uparrow}^{AP}+T_{\downarrow}^{AP})}{T_{\uparrow}^{AP}+T_{\downarrow}^{AP}}\label{eq:tmr},
\end{equation}
where the antiparallel (AP) configurations correspond to the $\ud$ and $\udu$ states for the bilayer and trilayer cases, respectively. Magnetoresistances based on the parallel configurations determined in this study (TMR$_{\uuu}$ $\approx$ 3.5 and TMR$_{\uparrow\uparrow}$ $\approx$ 1.7) are in good agreement with experiments (TMR$_{\uuu}$ $\approx$ 3.0 and TMR$_{\uparrow\uparrow}$ $\approx$ 1.0) \cite{klein2018}. A  difference in tunneling magnetoresistance (TMR$_{\uud}$ $\approx$ 0.7) is evident between the two antiparallel states with the same net magnetization, $\uud$ and $\udu$ (Fig.~\ref{fig:tmr2}), which we associate to the increase in conductance observed experimentally  at intermediate fields (TMR $\sim$0.5) \cite{klein2018}. Thereby, these variations in effective spin barriers may provide a method to probe the dynamics of metamagnetic configurations in the system during the switching process via external fields. Additionally, the TMR increase for energies above the Fermi level is due to an enhanced transmission of Bloch states near the conduction band edge of CrI$_3$. This phenomenon allows for the modulations of the TMR by a few orders of magnitude as recently demonstrated in dual-gated TMJ's \cite{wang2018very, song2018giant, song2019voltage}. Overall, weak layer-layer coupling plays a crucial role regarding electron transport prediction and intermediate state identification in these 2D layer channels.


\section{Conclusion}
In summary, we investigate ballistic spin transport through tunneling junctions consisting of few layer CrI$_3$ and graphene electrodes. Our obtained TMR values are in remarkable agreement with previous experimental reports.  Moreover, it shows that tunneling is the dominant transport mechanism, explaining disagreements found in previous calculations where high-strain supercells were employed. Interlayer coupling, despite usually being perceived as weak, is key to properly describing electronic properties and transport in these systems. The inclusion of these effects allows for the identification of intermediate metamagnetic states which are not captured by spin filter models that treat layers as independent. We also find that the tunneling rates in these short junctions vary considerably with respect to those estimated using the complex band structure, emphasizing the need to account for the atomistic details of the electrodes when quantifying this phenomenon in layered materials. Analysis of the energy dependence of tunneling rates provide mechanisms to enhance TMR and improve conductance values. The results of this work may prove valuable to the design and characterization of spin valves formed with 2D materials.

The authors would like to acknowledge the Auburn University Hopper HPC cluster for providing  resources which have contributed to the research results reported within this paper.
J.J.H. and M.A.K.~are thankful for startup funds at Auburn University and partial support from NSF DMR-1848344 grant.


\bibliography{./mybib}

\end{document}